\documentclass[aps,pre,twocolumn,groupedaddress, showkeys, a4paper]{revtex4}
\usepackage{amsmath}
\usepackage{graphicx}
\usepackage{hyperref}

\begin{document}


\title{Surprise maximization reveals the community structure of complex networks}


\author{Rodrigo Aldecoa}
\email[]{raldecoa@ibv.csic.es}
\author{Ignacio Mar\'in}
\email[]{imarin@ibv.csic.es}
\affiliation{Instituto de Biomedicina de Valencia.
Consejo Superior de Investigaciones Científicas (IBV-CSIC)
Calle Jaime Roig 11. Valencia, Spain}


\date{\today}

\begin{abstract}
How to determine the community structure of complex networks is an open question. It is critical to establish the best strategies for community detection in networks of unknown structure. Here, using standard synthetic benchmarks, we show that none of the algorithms hitherto developed for community structure characterization perform optimally. Significantly, evaluating the results according to their modularity, the most popular measure of the quality of a partition, systematically provides mistaken solutions. However, a novel quality function, called Surprise, can be used to elucidate which is the optimal division into communities. Consequently, we show that the best strategy to find the community structure of all the networks examined involves choosing among the solutions provided by multiple algorithms the one with the highest Surprise value. We conclude that Surprise maximization precisely reveals the community structure of complex networks.
\end{abstract}


\maketitle

\section*{}
The analysis of networks has profound implications in very different fields, from sociology to biology\cite{WassermanFaust1994, Strogatz2001, BarabasiOltvai2004, Costaetal2007, Newman2010}. One of the most interesting features of a network is its community structure\cite{Fortunato2010, Labatut2012}. Communities are groups of nodes that are more strongly or frequently connected among themselves than with the other nodes of the network. The best way to establish the communities present in a network is an open problem. Two related questions are still unsolved. First, which is the best algorithm to characterize networks of known community structure. Second, how to evaluate algorithm performance when the community structure is unknown. The first question requires testing the algorithms in benchmarks composed of complex networks where the community structure is established \emph{a priori}. In these benchmarks, it has been found that algorithm performance depends on how different is the density of intracommunity links from the average density of links in the network. In addition, it has been determined that most algorithms perform well when the networks are small and the communities have similar sizes, but many perform quite poorly in benchmarks composed of large networks with many communities of heterogeneous sizes\cite{GN2002, Danonetal2005, Danonetal2006, LFR, LFcomparison, Ormanetal2012, AldecoaMarin2011, RonhovdeNussinov2009, traag2011narrow, LF2011limits, AldecoaMarin2012}. Thus, benchmarks with the latter features have become crucial to rank algorithm performances. Among them, the Lancichinetti-Fortunato-Radicchi (LFR) benchmarks\cite{LFR, LFcomparison, Ormanetal2012, AldecoaMarin2011, RonhovdeNussinov2009, traag2011narrow, LF2011limits, AldecoaMarin2012} and the Relaxed Caveman (RC) benchmarks\cite{AldecoaMarin2011, Watts2003, AldecoaMarin2010} have shown to be particularly useful. Both benchmarks pose a stern test for algorithms that deal poorly with the presence of many communities, of small communities or of a mixture of communities of different sizes (see e. g. refs. \cite{LFR, Ormanetal2012, AldecoaMarin2011}).

The second question, how to determine the best performance when the community structure is unknown, involves devising an independent measure of the quality of a partition into communities that can be reliably applied to any type of network. The first and still today most popular such measure was called modularity\cite{NewmanGirvan2004} often abbreviated as Q). Modularity compares the number of links within each community with the expected number of links in a random graph of the same size and same distribution of node degrees and then adds the differences between expected and observed values for all the communities. It was proposed that the optimal partition of a network could be found by maximizing Q\cite{NewmanGirvan2004}. However, it was later determined that modularity-based evaluations are often incorrect when small communities are present in the network, i. e.  Q has a resolution limit\cite{FortunatoBarthelemy2007}. Several other works have found additional, subtle problems caused by using modularity maximization to determine network community structure\cite{LF2011limits, Goodetal2010, bagrow2012communities, XiangHu2012, Xiangetal2012}. All these results suggest that using Q provides incorrect answers in many cases.

We recently suggested an alternative global measure of performance, which we called Surprise\cite{AldecoaMarin2011}. Surprise assumes, as a null model, that links between nodes emerge randomly. It then evaluates the departure of the observed partition from the expected distribution of nodes and links into communities given that null model. To do so, it uses the following cumulative hypergeometric distribution:
\begin{equation}
S=\displaystyle-\log\sum\limits_{j=p}^{min(M,n)} \frac{\binom{M}{j}{\binom{F-M}{n-j}}}{\binom{F}{n}}
\end{equation}
Where $F$ is the maximum possible number of links in a network ($(k^2-k)/2$, being $k$ the number of units), $n$  is the observed number of links, $M$ is the maximum possible number of intracommunity links for a given partition, and $p$ is the total number of intracommunity links observed in that partition\cite{AldecoaMarin2011}. Using a cumulative hypergeometric distribution allows to exactly calculate the probability of the distribution of links and nodes in the communities defined for the network by a given partition. Thus, S measures how unlikely (or “surprising”, hence the name of the parameter) is the distribution of links and nodes in the communities defined in the network. In previous studies, we showed that Surprise improved on modularity in standard benchmarks and that choosing algorithms with high S values leads to accurate community structure characterization\cite{AldecoaMarin2011, AldecoaMarin2012}. Although these results were encouraging, whether S maximization could be used to obtain optimal partitions was not rigorously tested. This was due to the fact that Surprise values were estimated from the partitions provided by just a few algorithms. Given that other algorithms could provide even higher S values, it was unclear how optimal these results were.

Here, we test the best strategies currently available to characterize the structure of complex networks and we compare them with the results provided by Surprise maximization in both LFR and RC benchmarks. We first show that none among a large number of state-of-the-art algorithms work consistently well in all these complex benchmarks. Particularly, all modularity-based heuristics behave poorly. Also, we demonstrate that evaluating the performance of an algorithm using modularity is incorrect. We then show that a simple meta-algorithm, which consists in choosing in each network the algorithm that maximizes Surprise, very efficiently determines the community structure of all the networks tested. This method clearly performs better than any of the algorithms devised so far. We conclude that Surprise maximization is the strategy of choice for community characterization in complex networks.

\section*{Results}
\begin{figure*}[ht]
\includegraphics[scale=0.95]{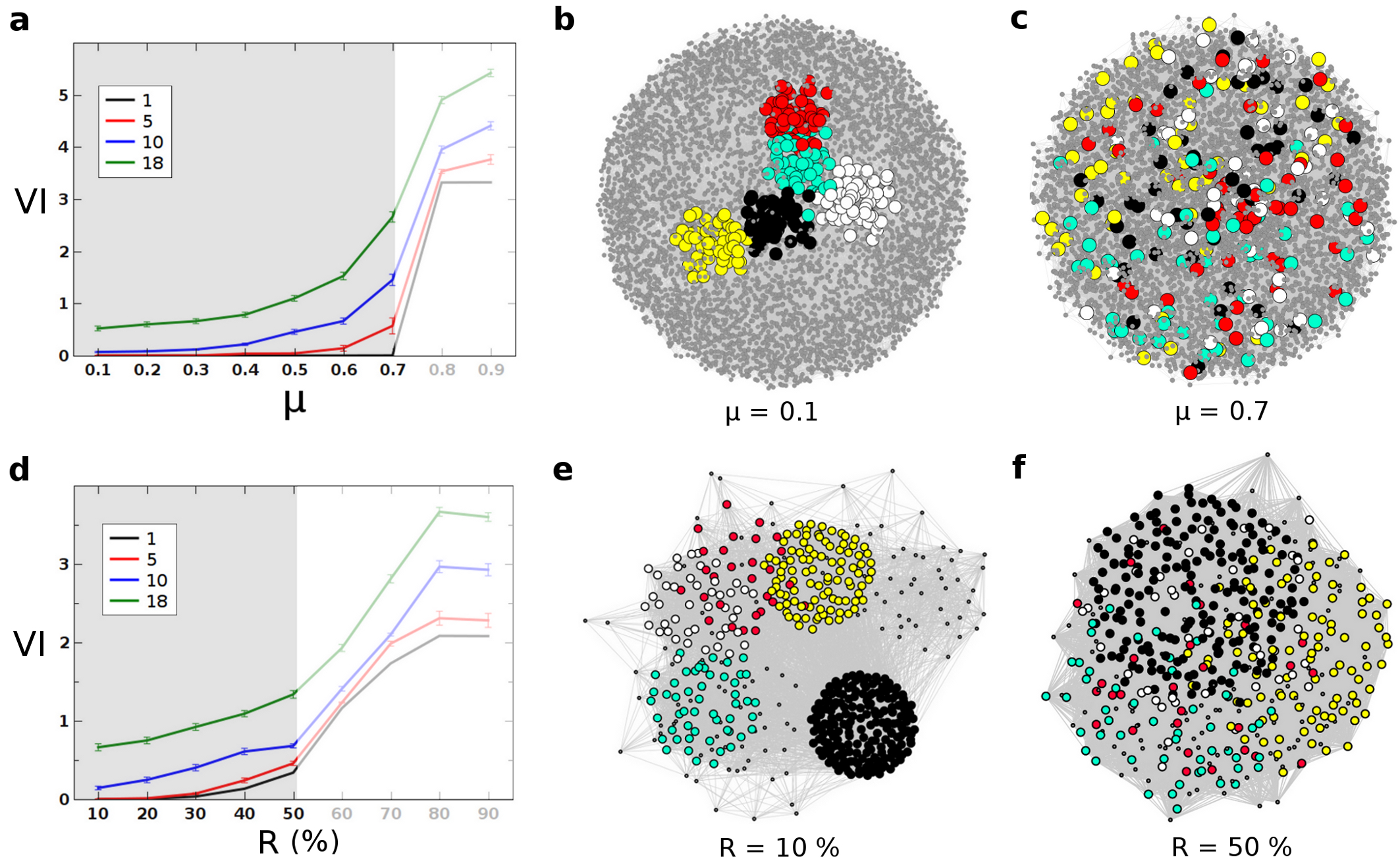}%
\caption{\label{fig:1} Global performance of the algorithms.
a) Behavior of the algorithms in the LFR benchmark. To obtain this figure, the algorithms were first ordered according to the VI results obtained for each $\mu$ condition. Then, we plotted the results for the algorithm with the best VI value (black line, indicated with "1"), the average of the top five algorithms (red line), average of the top ten ones (blue line) or average for all the 18 algorithms (green line). The grey region corresponds to the values of $\mu$ (0.1 - 0.7) chosen to perform the main comparative analyses (see text). Beyond that region, even the best algorithms obtain VI values considerably higher than zero, meaning that the original structure of the network has been significantly modified by the increase in intercommunity links. b) An example showing the five largest communities in a LFR network (5000 units) when $\mu=0.1$. Nodes are distributed into two dimensions with a spring-embedded algorithm\cite{LayoutAlgorithm} and drawn using Cytoscape\cite{Cytoscape}. Communities are well-isolated groups. c) The five largest communities when $\mu=0.7$. They are barely distinguishable in this representation because the mixing of links was quite extreme. However, several algorithms were still often able to detect these fuzzy communities. d) - f): The same results for the RC benchmark (512 units). Panel e depicts the five largest communities when R = 10 \% and Panel f to the same communities when R = 50 \%. Again, notice in panel d) the sharp increase in VI values when R $>$ 50 \%. An extreme degree of superimposition among communities is observed already when R = 50 \% (f). In the LFR benchmark, the rapid increase in VI values when the intercommunity links goes from $\mu$ = 0.7 to 0.8 (Panel a) is explained by all communities being of similar sizes. Therefore, they are destroyed at about the same time. On the contrary, the more progressive increase in VI when R grows, which we observed in Panel d, is due to the heterogeneous sizes of the communities present in that benchmark, which break down at different times.}
\end{figure*}

In order to determine the performance of different algorithms for community structure characterization, we explored two standard benchmarks, an LFR benchmark with 5000 units and an RC benchmark with 512 units (see Methods). Variation of Information (VI) was used to determine the degree of congruence between the partitions into communities suggested by 18 different algorithms and the real community structure present in the networks. A perfect congruence corresponds to a value VI = 0. Figures \ref{fig:1}a and \ref{fig:1}d display the general results obtained in the two benchmarks. 
A sharp VI increase was found when the community structure was weakened by highly increasing the number of intercommunity links, as occurs when the mixing parameter $\mu$ of the LFR benchmark has values above 0.7 or the rewiring parameter R of the RC benchmarks is higher than 50 \% (see also Methods for the precise definitions of $\mu$ These results mean that, above $\mu$ = 0.7 or R = 50 \%, the community structure originally present in the networks was substantially altered. In such cases, we could not determine whether the partitions suggested by the algorithms were correct or not: there would not be a known structure with which to compare. Thus, we decided to restrict our subsequent analyses to the LFR networks with $0.1 \le \mu \le 0.7$ (100 realizations per $\mu$ value, giving a total of 700 networks) and the RC networks with $10 \% \le R \le 50 \%$ (again, 100 realizations per R value, for a total of 500 different networks). These conditions generate some community structures that are very difficult to detect (Figure \ref{fig:1}).
\begin{figure}
\includegraphics[scale=0.55]{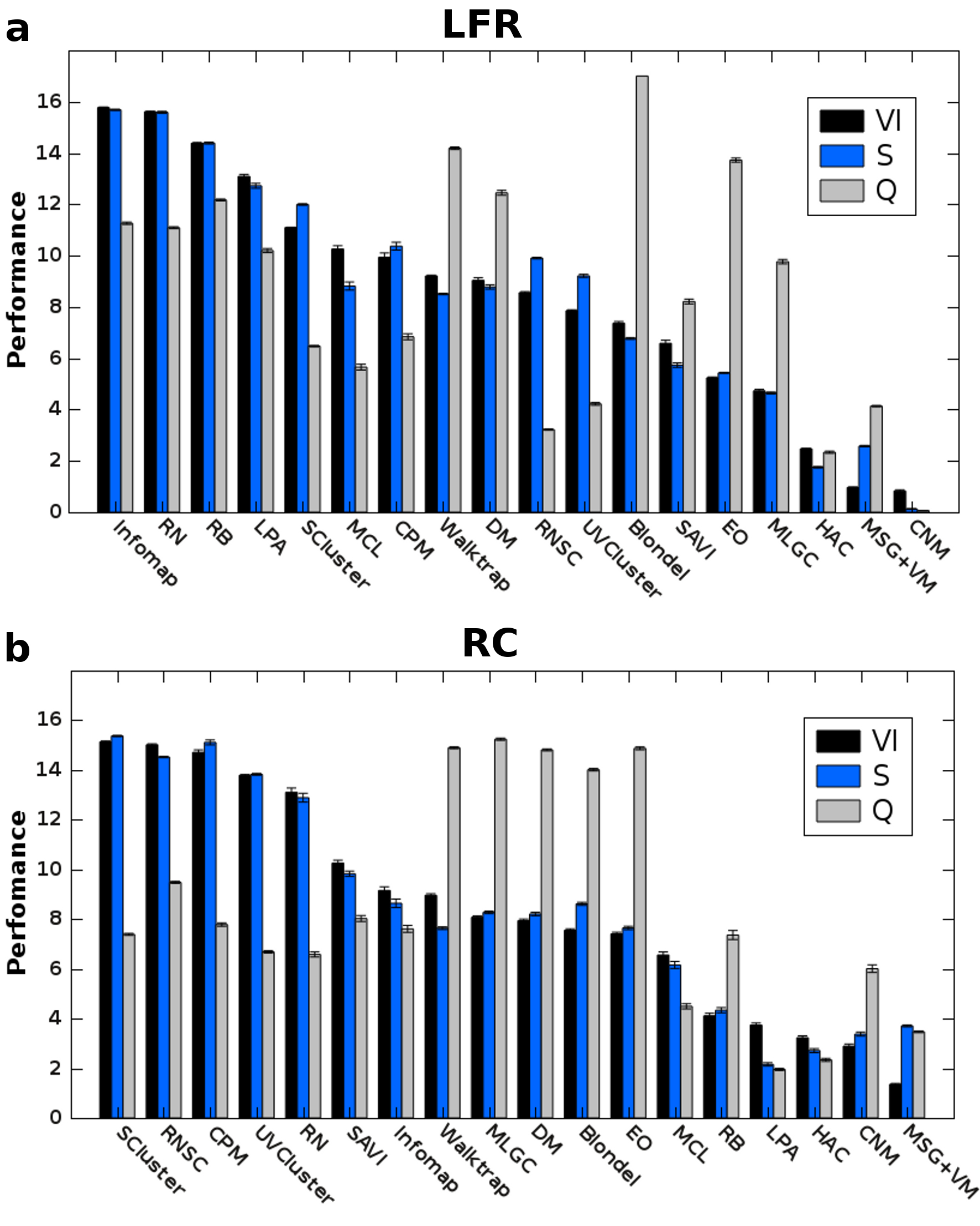}
\caption{\label{fig:2} Performance of the algorithms according to Variation of Information (VI), Surprise (S) and Modularity (Q) in LFR and RC benchmarks. Average performance and standard errors of the mean are shown. Performance values were obtained by the following method: 1) the VI, S or Q values of the partitions provided by the 18 algorithms in each of the networks (i. e. 700 values for LFR benchmarks, 500 values in RC benchmarks) were established; 2) For each network, the algorithms were assigned a rank according to their performance (1 = optimal, 18 = worse); identical ranks were given to tied algorithms (i. e. the ranks that would correspond to each of them were summed up and then divided by the number of tied algorithms); and, 3) Performance was calculated as 18 – average rank, meaning that 17 is the maximum possible value that would obtain an algorithm that outperforms the rest in all networks, and 0 equals to being the worst in all networks.}
\end{figure}

Figure \ref{fig:2} summarizes the individual performance of the algorithms according to three global measures of partition quality. The first one is VI, the gold standard for algorithm performance in these benchmarks. The other two, already mentioned above, are Surprise (S) and modularity (Q). The performance values measured according to the VI scores shown in Figure 2 indicate two very important facts. On one hand, none of the algorithms was the best in all LFR or in all RC networks. On the other hand, the best algorithms in LFR networks often performed poorly in RC networks, and vice versa (see e. g. the results of RB, LPA or RNSC in Figure 2). This can be rigorously shown by ordering within each benchmark the algorithms according to their performance, assigning a rank, from best to worst, and comparing the ranks in both benchmarks. We found that Kendall's non-parametric correlation coefficient for these ranks was very weak, just $\tau$ = 0.31 (p = 0.04, one-tailed comparison). We conclude that using single algorithms for community characterization is inadvisable, given that their performance is strongly dependent on the particular structure of the network.

If we focus now on the Surprise (S) and modularity (Q) results shown in Figure 2, another two striking facts become apparent. First, there was a very strong correlation between the performance of the algorithms according to VI and according to S. Kendall's correlation coefficient for the ranks of the performances of the algorithms ordered according to VI and to S values is $\tau$ = 0.91 in the LFR benchmarks (p = 4.9 x $10^{-11}$, one-tailed comparison) and $\tau$ = 0.83 in the RC benchmark (p = 1.4 x $10^{-8}$, one-tailed test). These results demonstrate that S is an excellent measure of the global quality of a division into communities, confirming and extending the conclusions of one of our previous works\cite{AldecoaMarin2011}. Second, the performance of the algorithms evaluated using Q only weakly correlated with their performance according to VI in the LFR benchmarks (Kendall's $\tau_{LFR}$ = 0.29, $p$ = 0.048, one-tailed test) and these two measures did not significantly correlate in the RC benchmarks ($\tau_{RC}$ = 0.27, $p$ = 0.66, again one-tailed test). These results indicate that evaluating the quality of a partition according to its modularity is inappropriate. It was therefore logical to find out that both the algorithms devised to maximize Q (Blondel, EO, MLGC, MSG+VM and CNM\cite{Blondel, EO, MLGC, MSG, CNM}) and the algorithms that use Q to evaluate the quality of their partitions (Walktrap, DM\cite{Walktrap, DM} were poor performers (Figure \ref{fig:2}).

\begin{figure}
\includegraphics[scale=0.55]{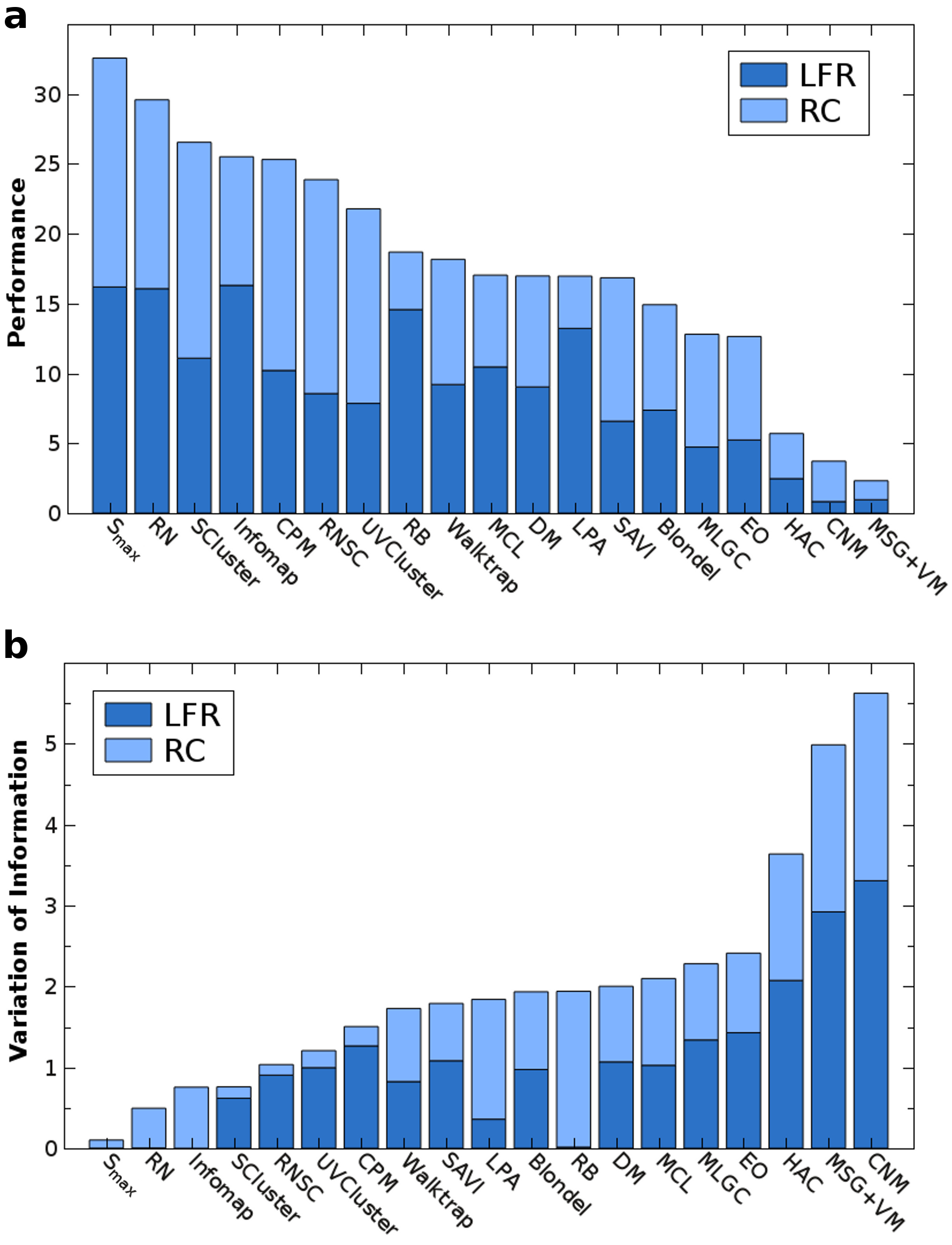}
\caption{\label{fig:3} A simple meta-algorithm based on Surprise maximization improves over all known community detection algorithms. a) Performances (calculated as in Figure \ref{fig:2}) for all the algorithms are compared in both the LFR and the RC benchmarks with the performance of a strategy that consists in picking up the algorithm that provides the highest S value ($S_{max}$). b) For the $S_{max}$ strategy, the average VI values for the 1200 networks analyzed are very close to zero, i. e. an almost optimal performance.}
\end{figure}

\begin{figure}
\includegraphics[scale=0.65]{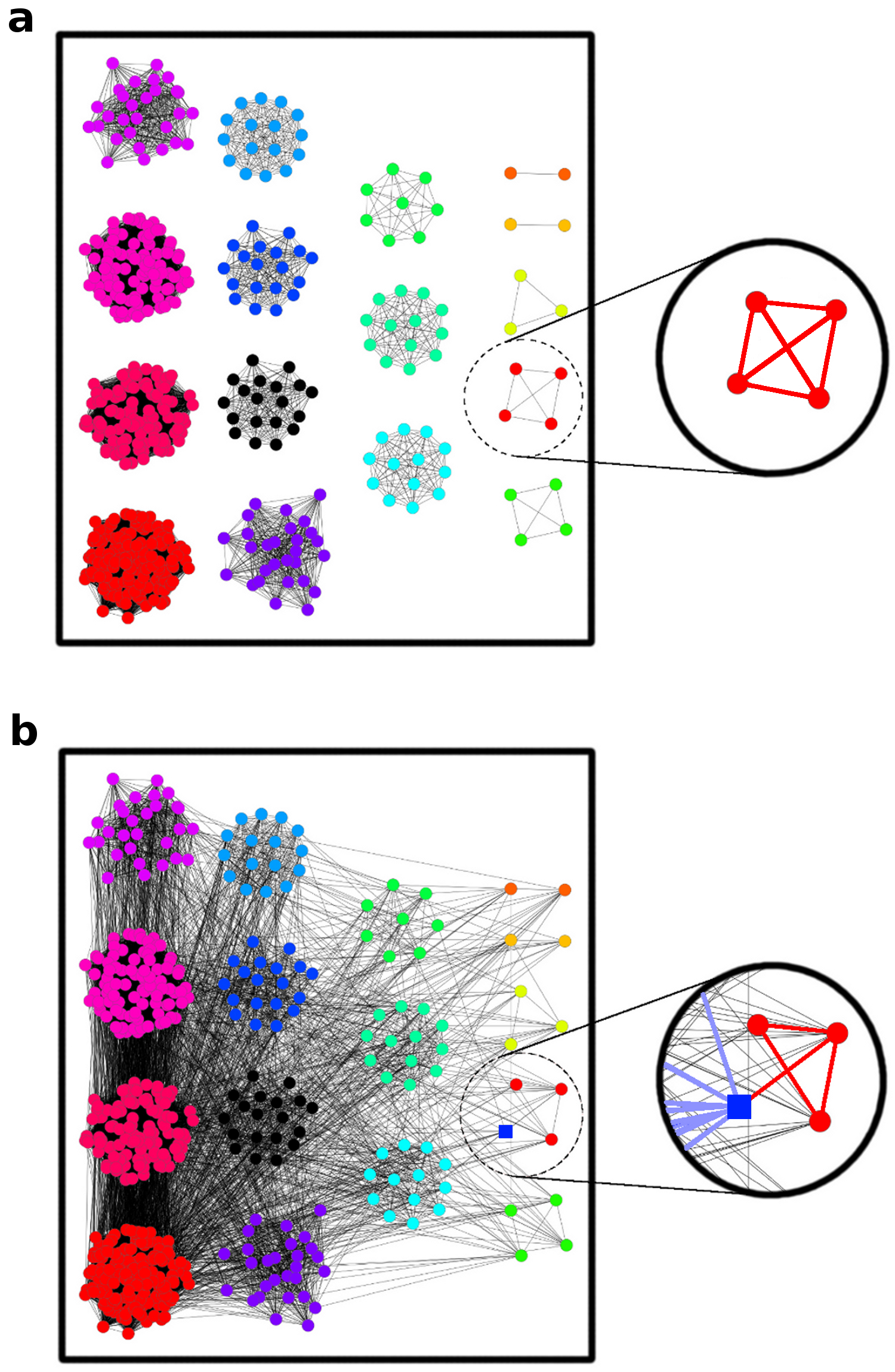}
\caption{\label{fig:4} When VI and Surprise maximum values do not coincide, the difference is often due to minimal changes in the community structure of the network. This is an example from the RC benchmark where $S_{max} > S_{orig}$ (see text). a) original structure. b) after R = 10 \% has been applied. $S_{max}$ is obtained when a single unit (square) is classified as being isolated from its original 4-nodes community (highlighted). As shown in panel b), the critical unit has become almost fully separated from the rest of the nodes in its original community, only one link remains, while it has been connected to many nodes in other communities.}
\end{figure}

If indeed maximization of Surprise is an optimal strategy for community characterization, as its strong correlation with VI suggests, then it should be possible to improve on the results of any single algorithm by simply picking up among many algorithms the one that generates the highest S value ($S_{max}$) in each particular network. Also, this S-maximization strategy should provide VI values very close to zero in our benchmarks. These two expectations are fulfilled, as shown in Figure \ref{fig:3}.
The top panel (Figure \ref{fig:3}a) demonstrates that choosing in each particular case the algorithm with the highest S value is better than selecting any of the state-of-the-art algorithms tested. It is remarkable that the $S_{max}$ values in Figure \ref{fig:3}a derived from the combined results of as many as 7 algorithms (CPM, Infomap, RB, RN, RNSC, SCluster and UVCluster\cite{traag2011narrow, AldecoaMarin2010, Infomap, RB, RN, RNSC, UVCluster}).  In addition, Figure \ref{fig:3}b indicates that the sum of the average VI values obtained using $S_{max}$ in the 1200 networks analyzed (with $\mu$ = 0.1 - 0.7 and R = 10 - 50 \%) were just slightly above zero, i. e. almost optimal. The average values were 0.002 $\pm$ 0.000 in the LFR benchmarks and 0.100 $\pm$ 0.007 in the RC benchmarks. We may ask why these VI values are not exactly zero, given that VI = 0 would be expected for a perfect global measure. We detected two reasons for this minor discrepancy. The first reason was that, in some cases (mainly in the RC benchmark with R = 50 \%), the available algorithms failed to obtain the highest possible S values. We found that the S values expected assuming that the original community structure of the network was intact ($S_{orig}$) were often higher than $S_{max}$  (Table \ref{table:1}). This obviously means that these algorithms did not found the community structure that maximizes S. That structure could still be the original one -- which indeed has the highest S value observed so far in our analyses -- or some alternate structure, but clearly not any of those found by the algorithms, which had lower S values. The second reason observed was the presence of minor changes in community structure that occurred in some networks when intercommunity links increased. Thus, the exact original structure of the network was not present anymore. This was deduced from the fact that $S_{max}$ values were sometimes slightly higher than $S_{orig}$ both in the LFR benchmarks with $\mu$ = 0.6 - 0.7 and the RC benchmarks with R = 10 - 40 \% (Table \ref{table:1}). These results suggested that the algorithms obtained optimal partitions, but they were a bit different from the original ones. To establish that fact, we examined the 23 cases where $S_{max} > S_{orig}$ in the RC benchmarks with R = 10 \%. We found that the partitions with $S_{max}$ values generally differed from the original structures in one of the smallest communities having lost single units (Supplementary Table 1; see example in Figure \ref{fig:4}).
Significantly, in those 23 networks we always found just one partition with $S_{max} > S_{orig}$ and several algorithms often recovered exactly that same partition (Supplementary Table 1). All these results indicate that real, small changes in community structure occurred in those networks, suggesting that the partitions with $S_{max} > S_{orig}$ values were indeed optimal. From the data in Table \ref{table:1}, we also obtain an indirect validation of our decision of using the LFR benchmarks with $\mu\le$ 0.7 and the RC benchmarks with R $\le$ 50 \% to evaluate algorithm performance. As shown in that Table, up to those limits, the $S_{max}$ and $S_{orig}$ values are not significantly different, while, beyond those limits, very significant differences are found. This means that the original structures, or structures almost identical to them, were indeed present in the networks examined to generate the results summarized in Figures \ref{fig:2} and \ref{fig:3}, which precisely was the only condition required for a reliable measure of algorithm performance.

The important results described in Figures 2 and 3 indicate that S maximization should allow determining with a very high precision the community structure of any network. We have explored whether this may be the case even when the community structure is very poorly defined by analyzing the results of our 600 additional networks, corresponding to the LFR benchmark with mixing parameter $\mu$ = 0.8 and $\mu$ = 0.9 (i. e. 200 networks) and the RC benchmark with R = 60 \% to R = 90 \% (400 networks). As indicated above, in these networks, the VI-based optimality criterion (i. e. VI = 0 means finding the original community structure) cannot be confidently used (Figure \ref{fig:1}; Table \ref{table:1}). However, alternative, unknown structures may be present that the algorithms should be able to detect. If this is the case, a reasonable prediction is that the algorithms that are providing the maximum S values in the conditions that are closest to those extreme ones (i. e. when $\mu$ = 0.7 in the LFR benchmarks and R = 50 \% in the RC benchmarks) should also provide the best S values in the most extreme networks.
\begin{figure}
\includegraphics[scale=0.55]{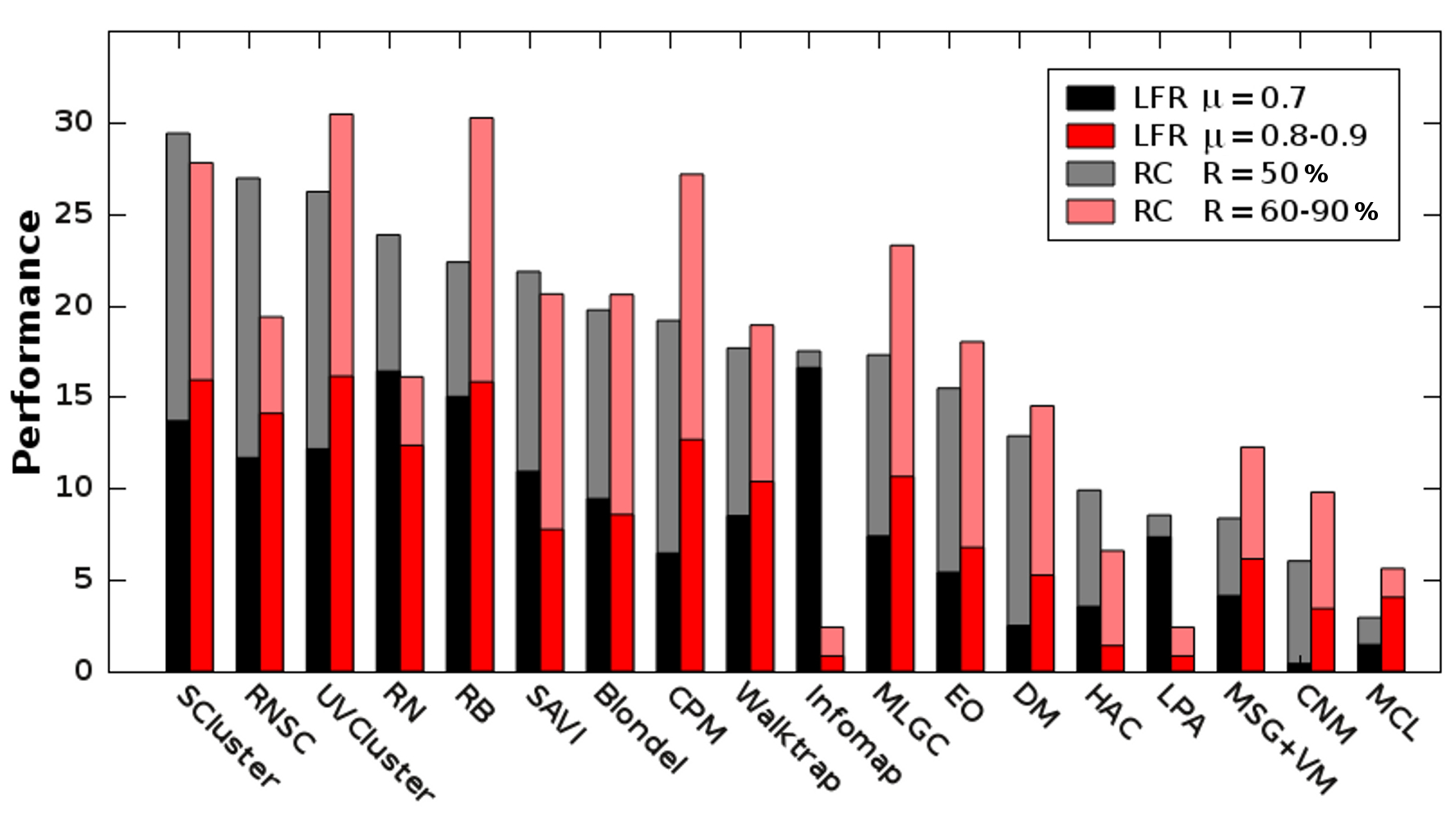}
\caption{\label{fig:5}The performance of the algorithms in the limit cases ($\mu$ = 0.7, R = 50 \%) and beyond those limits ($\mu$ = 0.8 - 0.9, R = 60 - 90 \%) are correlated. A statistically significant correlation was found, despite the fact that some algorithms, such as Infomap or LPA, totally collapsed. These algorithms established partitions consisting in a single community, which led to VI = 0 when compared with the original distribution.}
\end{figure}
Figure \ref{fig:5} shows that there is indeed a good correlation between the results obtained in the limit cases and in the most extreme cases. Kendall's non-parametric correlation coefficients for the ranks of the algorithms in the limit networks and in the most extreme networks are significant in both the LFR and RC benchmarks ($\tau_{LFR}$ = 0.42; $p$ = 0.007 and $\tau_{RC}$ = 0.49, $p$ = 0.020, one-tailed tests). This occurs despite some algorithms, as Infomap or LPA\cite{Infomap, LPA}, totally failing in these quasi-random networks (Figure \ref{fig:5}). UVCluster, RB, CPM and SCluster\cite{UVCluster, RB, traag2011narrow, AldecoaMarin2010} emerge as the best algorithms to characterize the structure in networks with poorly defined communities, in good agreement with previous results\cite{AldecoaMarin2011,traag2011narrow}.

\begin{figure}
\includegraphics[scale=0.8]{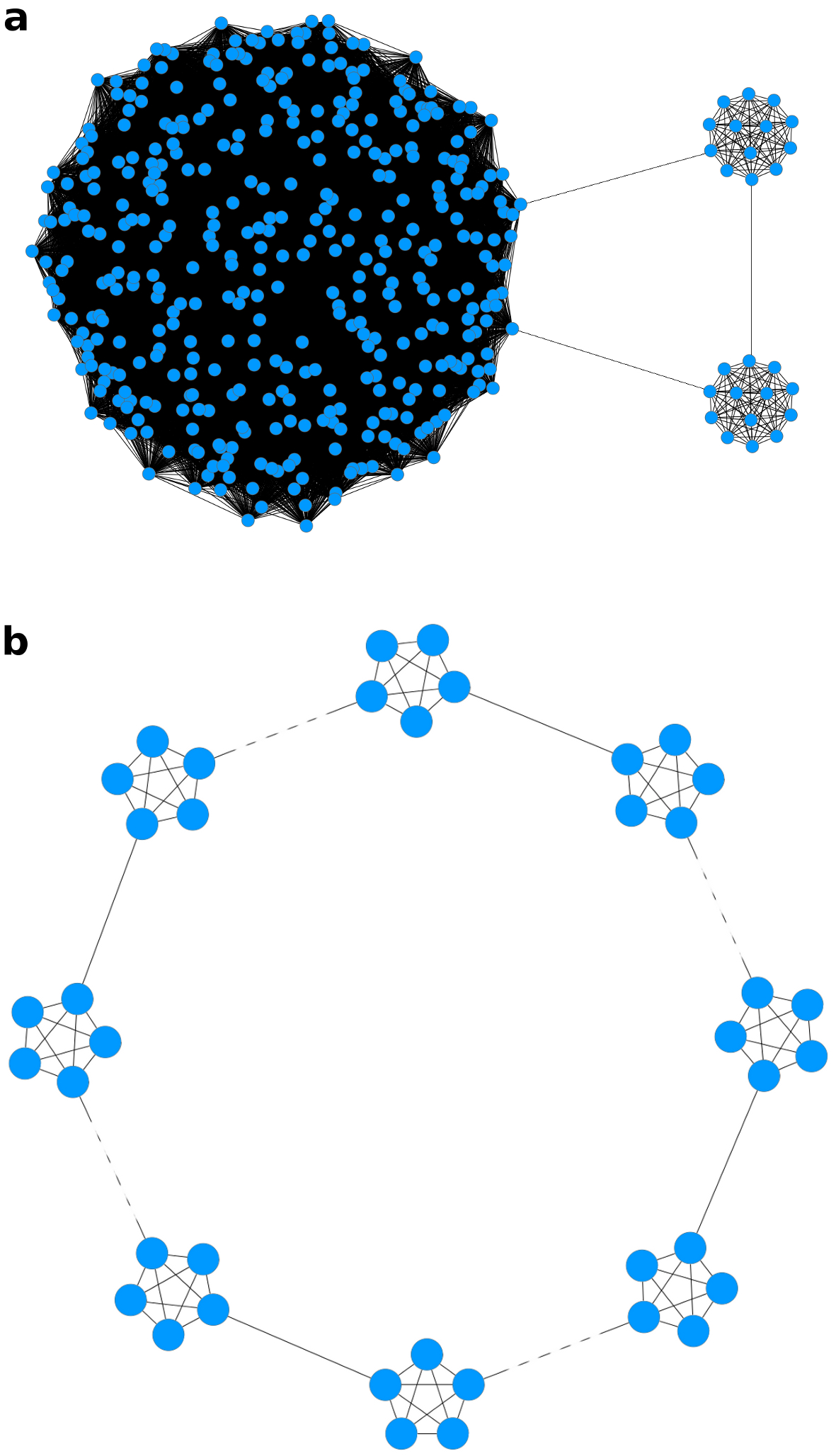}
\caption{\label{fig:6} Two extreme networks designed to test the behavior of Surprise when small communities are present. a) A network with three communities (sizes 400, 13, 13). The nodes of the largest community have an average degree of 100, while the nodes in the two smallest communities form cliques. The three communities are interconnected by single links, as shown. b) Cliques, each one with five nodes, which are connected also by single links in a way that can be depicted as a ring. The figure shows an example with eight cliques, but that number was progressively increased to determine whether the partition with highest S still corresponded to the one in which each cliques was an independent community.}
\end{figure}

We decided to perform some final tests to determine whether the limitations that affect Q when communities are very small may also affect S. For this purpose, we used two extreme networks of known structure suggested before\cite{LF2011limits, FortunatoBarthelemy2007, Granelletal2012}. The first one includes just three communities, one of them very large (400 nodes with average degree = 100) and the other two much smaller (cliques of 13 nodes). These three communities are connected by single links (Figure 6a). We found in this network that the maximum value of Q (EO algorithm, Q = 0.0836) did not correspond to a partition into the three natural communities. On the contrary, and as already noted by other authors in similar cases\cite{LF2011limits, FortunatoBarthelemy2007}, Q indicates a mistaken solution, in this case with five communities. On the other hand, the three communities were correctly found by multiple algorithms (CPM, Infomap, LPA, RNSC, SCluster and UVCluster), and this partition indeed corresponded to the maximum value of S (1230.73). The second extreme type of network was precisely the ring of cliques in which the resolution limit of Q was first described\cite{FortunatoBarthelemy2007}, which is schematized in Figure 6b.  Here, a variable number of cliques, each one composed of five units, were connected to each other by single links to form a ring. We were interested in determining whether, even if we increase the number of cliques, a solution in which all cliques are separately detected always has a better S value than one in which pairs of cliques are put together. We tested networks of sizes up to 1 million nodes, finding that the best partition was always the one in which the cliques are considered independent communities. On the contrary, when Q is used, the cliques are considered independent units only if the network size is smaller than 150 units.

\section*{Discussion}
Our results lead to two main conclusions. The first one is that none of the algorithms currently available generates optimal solutions in all networks (Figures 2, 3). In fact, there is just a weak correlation of the algorithm performances in the two standard benchmarks used in this study. More precisely, we can say that there are some algorithms that clearly fail in both benchmarks and the rest tend to perform much better in one of the benchmarks than in the other (see Figure 2). Most of the best overall performers were already found to be outstanding in other studies\cite{LFcomparison, Ormanetal2012, AldecoaMarin2011, traag2011narrow, AldecoaMarin2012, RN}. The exception is RNSC\cite{RNSC}, which had not been tested in depth before. Among the ones that always perform poorly are all the algorithms that use modularity as either a global parameter to maximize or as a way to evaluate partitions. This fact, together with the demonstration that Q does not correlate with VI in networks of known structure (Figure 2), and also the good performance of S, including its ability to cope with extreme networks in which Q traditionally fails due to its resolution limit (Figure 6), should definitely deter researchers from using modularity. A strong corollary is that it is advisable a reevaluation of the hundreds of papers -- in fields as varied as sociology, ecology, molecular biology or medicine -- which are based on modularity analyses.

The second, and most important, conclusion is that the community structure of a network can be determined by maximizing S, for example by simply taking the results of as many algorithms as possible and choosing the one that provides partitions with the highest Surprise value. In a previous paper, we showed that Surprise can be used to efficiently evaluate the quality of a partition, behaving much better than modularity\cite{AldecoaMarin2011}, but the precise performance of the S-maximization strategy was not determined. Here, we extend those results, to show that using S maximization leads to an almost perfect performance. We were very close to solve the correct community structure of all the networks of these two benchmarks, as is strikingly demonstrated by the $S_{max}$ results shown in Figure 3b. It is significant that they were obtained by combining results of the 7 algorithms with the best average performances, as detailed in Figure 3a: RN, SCluster, Infomap, CPM, RNSC, UVCluster and RB. Another important result is summarized in Figure 5, which indicates that Surprise can also be used in cases in which the community structure is so blurred as to become almost random. Given these results, we conclude that Surprise is the parameter of choice to characterize the community structure of complex networks. Future works should use Surprise maximization, instead of modularity maximization or other methods, to establish that structure.

It is significant that only two algorithms (SCluster and UVCluster) use the maximization of Surprise to choose among partitions generated by consensus hierarchical clustering\cite{AldecoaMarin2010, UVCluster}. This may explain their good average results (Figures 2, 3, 5). However, no available algorithm performs searches to directly determine the maximal Surprise values. That type of algorithms could overcome the limitations detected in all the currently available ones, potentially allowing the characterization of optimal partitions even in the most difficult networks.

We may ask why Surprise is able to evaluate with such efficiency the quality of a given partition, while modularity cannot. In our opinion, the difference rests on the fact that modularity is based on an inappropriate definition of community. Newman and Girvan\cite{NewmanGirvan2004} verbally defined a community as a region of a network in which the density of links is higher than expected by chance. However, the precise mathematical model used to deduce the modularity formula implies a definition of community that does not take into account the number of nodes required to achieve such a high density\cite{NewmanGirvan2004}. By not evaluating the number of nodes, modularity falls prey of a resolution limit: small communities cannot be detected\cite{LF2011limits, FortunatoBarthelemy2007}. On the other hand, Surprise analyses often choose as best a solution where some communities are just isolated units (see examples in Figure 4 and Ref. 14). This happens because the Surprise formula precisely evaluates not only the number of links, but also the number of nodes within each community. For instance, incorporating a single poorly connected unit into a community is often forbidden by the fact that such incorporation sharply increases the number of potential intracommunity links (all those that might connect the units already present in the community with the new unit) while barely increasing the number of real intracommunity links. This leads to an S value much smaller than if the unit is kept separated. It is also significant that a general problem of modularity maximization and other related algorithms - as those based on Potts models with multiresolution parameters - is that they cannot find a perfect equilibrium between merging and splitting communities\cite{LF2011limits, XiangHu2012, Xiangetal2012}. In these methods, each community is evaluated independently, one at a time. The global value to be maximized is the sum of the qualities of the individual communities. However, in complex networks with communities of very different sizes, it may be often impossible to find a single rule (even using a tunable parameter, as in these multiresolution methods) to split some communities while keeping intact the rest17. Surprise analyses are not affected by this problem, because communities are not defined independently, one by one, but emerge as regions of nodes statistically enriched in links, according to the general features (i. e. the total number of nodes and links) of the whole network.

\section*{Methods}
We searched the literature to select the best community detection algorithms available to analyze networks with unweighted, undirected links. Our final results are based on 18 of them (summarized in Table \ref{table:2}). Algorithms known to behave poorly in similar benchmarks or specifically designed to characterize communities with overlapping nodes were discarded.  Some other algorithms that seemed interesting but we were unable to test for diverse reasons (e. g. they were not provided by the authors, did not complete the benchmarks, etc.) are detailed in Supplementary Table 2. We performed extensive tests with these selected algorithms, using their default parameters, in two very different benchmarks. They were chosen both difficult and very dissimilar, with the idea that the results could be general enough as to be extrapolated to networks of unknown structure. The first was a standard LFR benchmark already used in other studies that compared algorithms\cite{LFcomparison, Ormanetal2012, AldecoaMarin2011, LF2011limits}. It is composed of networks with 5000 nodes, structured in small communities with 10-50 nodes. The distribution of node degrees and community sizes were generated according to power laws with exponents -2 and -1, respectively. The sizes of the communities in the networks of this benchmark have average Pielou's indexes\cite{Pielou} with a value of 0.98. This index is equal to 1 when all communities are of the same size. The chief difficulty of this benchmark thus lies on the presence of many small communities. The second benchmark was one of the Relaxed Caveman (RC) type, very similar to the ones used in our previous works\cite{AldecoaMarin2011, AldecoaMarin2012}. The networks in this RC benchmark have 512 units and 16 communities, with sizes defined according to a broken-stick model to obtain an average Pielou's index = 0.75. This makes this benchmark very difficult, given that it consists of networks with communities of very different sizes, some of them very small (see e. g. Figure 4).  It was not convenient to our purposes to use larger RC benchmarks given that the total number of links in these networks quickly grows when the number of nodes is increased and many algorithms become too slow. 

These two benchmarks are "open", meaning that they have a tunable parameter that, when increased, makes the network community structure to become less and less obvious until it shifts towards a totally unknown structure, potentially very different from the original one and close to random\cite{LFR, Ormanetal2012, AldecoaMarin2011, AldecoaMarin2012}. This parameter increases intracommunity links and lowers the number of intercommunity links. In the case of the LFR benchmarks, the “mixing parameter”, $\mu$, indicates the fraction of links connecting each node of a community with nodes outside of the community\cite{LFR}. For the RC benchmarks, we defined Rewiring (R) as the percentage of links that is randomly shuffled among units. Thus, R = 10 \% means that 10 per cent of the links were first randomly removed and then added again, to link randomly chosen nodes.

Variation of information (VI)\cite{VI} was used to measure the agreement between the original community structure present in the network and the structure deduced by each algorithm. The advantages of using VI have been discussed in our previous works\citep{AldecoaMarin2011, AldecoaMarin2012}. A perfect agreement with a known structure will provide a value of VI = 0. In addition, two global quality functions, Newman and Girvan’s modularity (Q)\cite{NewmanGirvan2004} and Surprise (S)\cite{AldecoaMarin2011}, (see Formula [1]), were also used to evaluate the results. The values of S and Q for the partitions proposed by each algorithm were calculated and then all the values were used to determine the correlations of S and Q with VI and to establish these maximum values of S.

\begin{table*}[p]
	\begin{center}
    \begin{tabular}{|c||c|c|c|}
        \hline
		\multicolumn{4}{ |c| }{\textbf{LFR benchmark}} \\ \hline
        $\mu$ & $S_{orig}$ & $S_{max}$ & $p$        \\ \hline
        0.1 & 99065.69 $\pm$ 111.50 & 99065.69 $\pm$ 111.50 & ns       \\ 
        0.2 & 82631.18 $\pm$ 93.92  & 82631.18 $\pm$ 93.92  & ns       \\ 
        0.3 & 67847.35 $\pm$ 90.78  & 67847.35 $\pm$ 90.78  & ns       \\ 
        0.4 & 54354.47 $\pm$ 76.71  & 54354.47 $\pm$ 76.71  & ns       \\ 
        0.5 & 41991.16 $\pm$ 48.70  & 41991.16 $\pm$ 48.70  & ns       \\ 
        0.6 & 30807.18 $\pm$ 40.09  & 30807.38 $\pm$ 40.09  & ns       \\ 
        0.7 & 20563.37 $\pm$ 26.92  & 20570.70 $\pm$ 26.78  & ns       \\ 
        \textit{0.8} & \textit{11598.83} $\pm$ \textit{17.91}  & \textit{10168.11} $\pm$ \textit{28.15}  & $<$ \textit{0.0001} \\ 
        \textit{0.9} & \textit{4204.50} $\pm$ \textit{7.62}    & \textit{8368.94} $\pm$ \textit{4.21}    & $<$ \textit{0.0001} \\ \hline
   		\multicolumn{4}{}{} \\ \hline
		\multicolumn{4}{ |c| }{\textbf{RC benchmark}} \\ \hline
        $R$  & $S_{orig}$            & $S_{max}$             & $p$        \\ \hline
        10 & 19012.72 $\pm$ 67.33 & 19012.94 $\pm$ 67.32 & ns       \\ 
        20 & 13505.84 $\pm$ 34.14 & 13506.72 $\pm$ 34.11 & ns       \\ 
        30 & 9298.98 $\pm$ 11.88  & 9301.12 $\pm$ 11.88  & ns       \\ 
        40 & 6013.69 $\pm$ 3.92   & 6017.58 $\pm$ 4.09   & ns       \\ 
        50 & 3487.65 $\pm$ 11.54  & 3479.92 $\pm$ 12.99  & ns       \\ 
        \textit{60} & \textit{1647.42} $\pm$ \textit{13.82}  & \textit{1540.76} $\pm$ \textit{16.79}  & $<$ \textit{0.0001} \\ 
        \textit{70} & \textit{475.35} $\pm$ \textit{10.42}   & \textit{899.96} $\pm$ \textit{7.98}    & $<$ \textit{0.0001} \\ 
        \textit{80} & \textit{11.84} $\pm$ \textit{1.52}     & \textit{963.73} $\pm$ \textit{9.42}    & $<$ \textit{0.0001} \\ 
        \textit{90} & \textit{0.00} $\pm$ \textit{0.00}      & \textit{1003.21} $\pm$ \textit{9.95}   & $<$ \textit{0.0001} \\
        \hline
    \end{tabular}
	\end{center}
	\caption{\label{table:1}Average $S_{orig}$ and $S_{max}$ values in the LFR and RC benchmarks. Statistical significance ($p$) was estimated using a two-tailed Student t test. ns: non-significant differences. In italics, the benchmarks containing quasi-random networks, discarded for the main analyses (summarized in Figures \ref{fig:2} and \ref{fig:3}), but included in the analyses shown in Figure \ref{fig:5}}
\end{table*}

\begin{table*}
\begin{center}
    \begin{tabular}{|c|c|c|}
        \hline
        \textbf{Name of the Algorithm} & \textbf{Strategy used by the algorithm}              & \textbf{References} \\ \hline
        Blondel               & Multilevel modularity maximization          & \cite{Blondel}       \\ 
        CNM                   & Greedy modularity maximization              & \cite{CNM}       \\ 
        CPM                   & Multiresolution Potts model                 & \cite{traag2011narrow}       \\ 
        DM                    & Spectral analysis + modularity maximization & \cite{DM}       \\ 
        EO                    & Modularity maximization                     & \cite{EO}       \\ 
        HAC                   & Maximum Likelihood                          & \cite{HAC}       \\ 
        Infomap               & Information compression                     & \cite{Infomap}       \\ 
        LPA                   & Label propagation                           & \cite{LPA}       \\ 
        MCL                   & Simulated flow                              & \cite{MCL}       \\ 
        MLGC                  & Multilevel modularity maximization          & \cite{MLGC}       \\ 
        MSG+VM                & Greedy modularity maximization + refinement & \cite{MSG}       \\ 
        RB                    & Multiresolution Potts model                 & \cite{RB}       \\ 
        RN                    & Multiresolution Potts model                 & \cite{RN}       \\
		RNSC				  & Neighborhood tabu search                    & \cite{RNSC}		 \\
        SAVI      			  & Optimal prediction for random walks         & \cite{SAVI}     \\
        SCluster  			  & Hierarchical Clustering + Surprise maximization & \cite{AldecoaMarin2010}     \\ 
        UVCluster 		      & Hierarchical Clustering + Surprise maximization & \cite{UVCluster, AldecoaMarin2010} \\ 
        Walktrap              & Random walks + modularity maximization          & \cite{Walktrap}     \\ 
        \hline
    \end{tabular}
\end{center}
\caption{\label{table:2}Details of the algorithms used in this study. A summary of the strategies implemented by the algorithms and the corresponding references are indicated.}
\end{table*}

\pagebreak
\bibliography{Smax}

\end{document}